\documentstyle[aaspp4,emulateapj5,apjfonts]{article}

\gdef\hdfs{HDF-S}

\gdef\h50min{$h_{50}^{-1}$}
\gdef\kms{km\,s$^{-1}$}

\gdef\3727{[O\,{\sc ii}]\,3727\,\AA}
\gdef\5007{[O\,{\sc iii}]\,5007\,\AA}

\gdef\ms1054{MS\,1054--03}
\gdef\4ang{4000\,\AA}
\lefthead{van Dokkum et al.}
\righthead{Red Galaxies at $z\gtrsim 2$}
\slugcomment{Accepted for publication in Astrophysical Journal Letters}
\begin{document}

\title{Spectroscopic Confirmation of a Substantial
Population of Luminous Red
Galaxies at Redshifts $z \gtrsim 2$
\altaffilmark{1,2}}

\author{Pieter~G.~van Dokkum\altaffilmark{3,4}, 
Natascha M.~F\"orster Schreiber\altaffilmark{5},
Marijn Franx\altaffilmark{5},
Emanuele Daddi\altaffilmark{6},
Garth D.~Illingworth\altaffilmark{7},
Ivo Labb\'e\altaffilmark{5},
Alan Moorwood\altaffilmark{6},
Hans-Walter Rix\altaffilmark{8},
Huub R\"ottgering\altaffilmark{5},
Gregory Rudnick\altaffilmark{9},
Arjen van der Wel\altaffilmark{5},
Paul van der Werf\altaffilmark{5},
and Lottie van Starkenburg\altaffilmark{5}
}

\altaffiltext{1}
{Based on observations collected at the European Southern Observatory,
Paranal, Chile (ESO LP 164.O-0612)}
\altaffiltext{2}
{Based on observations obtained at the W.\ M.\ Keck Observatory,
which is operated jointly by the California Institute of
Technology and the University of California.}
\altaffiltext{3}{Department of Astronomy, Yale
University, New Haven, CT 06520-8101}
\altaffiltext{4}{California Institute of Technology, MS105-24, Pasadena,
CA 91125}
\altaffiltext{5}{Leiden Observatory, P.O. Box 9513, NL-2300 RA, Leiden,
The Netherlands}
\altaffiltext{6}{European Southern Observatory, Karl-Schwarzschild-Str.~2,
D-85748, Garching, Germany}
\altaffiltext{7}{UCO/Lick Observatory, University of California, Santa
Cruz, CA 95064}
\altaffiltext{8}{Max-Plank-Institut f\"ur Astronomie, K\"onigstuhl 17,
Heidelberg, Germany}
\altaffiltext{9}{Max-Plank-Institut f\"ur Astronomie, Karl-Scharzschild-Str.~1,
Garching, Germany}

\begin{abstract}

We confirm spectroscopically
the existence of a population of galaxies
at $z\gtrsim 2$ with rest-frame optical colors similar to normal nearby
galaxies. The galaxies were identified
by their red near-infrared colors
in deep images
obtained with ISAAC on the Very Large Telescope
of the field around the foreground cluster \ms1054.
Redshifts of six galaxies with $J_s -K_s>2.3$
were measured from optical spectra obtained with the W.\ M.\ Keck
Telescope. Five out of six
are in the range $2.43 \leq z \leq 3.52$, demonstrating
that the $J_s-K_s$ color selection is quite efficient.
The rest-frame ultraviolet spectra of confirmed $z>2$ galaxies
display a range of properties, with two galaxies showing emission
lines characteristic of AGN, two having
Ly$\alpha$ in emission, and one
showing interstellar absorption lines only.
Their full spectral energy distributions
are well described
by constant star formation models with
ages $1.4-2.6$\,Gyr, except for one galaxy whose colors indicate
a dusty starburst.
The confirmed $z>2$
galaxies are very luminous: their $K_s$
magnitudes are in the range $19.2-19.9$, corresponding
to rest-frame absolute $V$ magnitudes $-24.8$ to $-23.2$.
Assuming that our
bright spectroscopic sample is representative for the general
population of $J_s-K_s$ selected objects,
we find that the surface density of red $z \gtrsim 2$
galaxies is $\approx 0.9$~arcmin$^{-2}$ to $K_s=21$.
The surface density is comparable to
that of Lyman-break selected galaxies with $K_s<21$,
when corrections are made
for the different redshift distributions
of the two samples. Although there will be some overlap
between the two populations,
most ``optical-break'' galaxies are too
faint in the rest-frame ultraviolet to be selected as
Lyman-break galaxies.
The most straightforward
interpretation is that star formation
in typical optical-break galaxies
started earlier than in typical Lyman-break
galaxies. Optical-break galaxies may be the oldest and most
massive galaxies yet identified at $z> 2$, and
could evolve into early-type galaxies and bulges.

\end{abstract}



\section{Introduction}

The identification of star-forming galaxies at $z \gtrsim 3$ by the
Lyman-break technique has greatly enhanced our understanding of galaxy
formation and the star formation history of the universe ({Steidel}
{et~al.} 1996, 1999; {Madau} {et~al.} 1996).  Although other high
redshift galaxy populations have since been identified (e.g., {Hu},
{Cowie}, \& {McMahon} 1998; {Smail} {et~al.} 2000; {Barger} {et~al.}
2001), Lyman-break galaxies (LBGs) dominate the UV luminosity density
at high redshift ({Steidel} {et~al.} 1999), and have been argued to be
the progenitors of massive galaxies in groups and clusters (e.g.,
Baugh et al.\ 1998).

However, the census of normal galaxies at $z \sim 3$ may still be
incomplete because of selection effects. The Lyman-break technique
requires a high rest-frame far-ultraviolet (UV) luminosity, and so
will preferentially select relatively unobscured, actively
star-forming galaxies. Stellar ages of LBGs are typically $\sim 3
\times 10^8$\,yr ({Papovich}, {Dickinson}, \& {Ferguson} 2001;
{Shapley} {et~al.} 2001), suggesting that the descendants of galaxies
which started forming stars at significantly higher redshift ($z> 4$)
may be underrepresented in current surveys (e.g., {Ferguson},
{Dickinson}, \& {Papovich} 2002).

As shown in a companion paper (Franx et al.\ 2003) 
such ``evolved'' high redshift galaxies can be
selected effectively in the rest-frame optical, which is redshifted
to near-infrared (NIR) wavelengths for $z\gtrsim 2$.  The 
criterion $J_s-K_s>2.3$
is expected to select galaxies with
prominent rest-frame optical breaks,
caused by the 3625\,\AA\ Balmer-break or
the 4000\,\AA\ Ca\,{\sc II} H+K break.\footnote{Magnitudes are
on the Vega system, unless stated otherwise.}
This ``optical-break'' selection
is complementary to the
ultraviolet Lyman-break selection.
In Franx et al.\ (2003) we show that optical-break galaxies
have a high surface density in Hubble Deep Field South (\hdfs), and derive
from photometric redshifts  that their volume density is
comparable to LBGs.  In Daddi et al.\ (2003) we infer that
the population is highly clustered.
The main uncertainties in Franx et al.\
are the lack of spectroscopic redshifts and the small $2\farcm 3
\times 2\farcm3$ field.

Here we present
spectroscopic observations of a small sample of
optical-break galaxies in \ms1054, a field with a foreground cluster
at $z=0.83$.  The NIR imaging data in this field complement
our extremely deep Very Large Telescope (VLT) ISAAC data on
\hdfs\ (Labb\'e et al.\ 2003), and consist of
77.5 hours of $J_s,H,K_s$ imaging with ISAAC
distributed over a $5\farcm 4
\times 5\farcm 4$ square of four pointings
(F\"orster Schreiber et al., in preparation).  Although not as deep as
our \hdfs\ observations, the $\sim 5$ times larger area
enables a more robust measurement of the surface density and is well
suited for efficient multi-slit spectroscopy.

\section{Spectroscopy}

Optical spectroscopy was obtained for galaxies in the
\ms1054\ field on
2002 February 14--17 with the Low Resolution Imaging Spectrograph
(Oke et al.\ 1995) on the W.\ M.\ Keck telescope. The sample was not
restricted to $J_s-K_s$ selected galaxies, but also contained candidate
$1 < z < 2$ galaxies selected by their red $I - H$ color.
The sample was limited at
$K = 20.7$, with priority given to objects with $I < 24.5$.
Four multi-slit masks were designed, with faint objects repeated
in several or all masks. The 300 lines\,mm$^{-1}$ grism was used on the
blue arm, in combination with the D680 dichroic. On the
red arm we used the 600 lines\,mm$^{-1}$
grating (1\,$\mu$m blaze) on February 14-15 and
the 400 lines\,mm$^{-1}$ grating (8500\,\AA\ blaze)
on February 16--17. Conditions were photometric and the seeing
ranged from $0\farcs 8$ to $1\farcs 5$.
The exposure time ranged from 7.2\,ks for the
brightest objects to 72\,ks for the faintest.

Redshifts were measured 
for six out of eleven observed $J_s - K_s$ selected galaxies.
Five have redshifts in the range
$2.43- 3.52$, with the remaining galaxy showing a faint
emission line which we tentatively identify as \3727\
at $z = 1.19$ (Table 1). Additionally, redshifts were measured for twelve out
of 27 observed galaxies with $I - H > 3.0$ and $J_s - K_s < 2.3$;
they are in the range $1.06 - 1.87$, with mean $\langle z
\rangle = 1.40$.

The spectroscopic redshifts confirm
that our simple $J_s - K_s$ color cut effectively isolates
galaxies at $z>2$, indicating only a small
($\sim 20$\,\%) contamination by (dusty)
galaxies at lower redshifts.
A concern is that the five $J_s-K_s$ selected galaxies without
redshift are at $z \sim 1.5$, as there would be no strong
spectral features in the observed wavelength range.
The importance of this potential bias can be estimated by
comparing the photometric redshifts of galaxies
with and without spectroscopic redshift.
Photometric redshifts were determined
using the publically available {\sc hyperz} code
(Bolzonella, Miralles, \& Pell\'o 2000); a full discussion is given
in F\"orster Schreiber et al., in preparation.
As demonstrated in Fig.\ \ref{zdist.plot} the photometric and
spectroscopic redshifts are in good agreement. The red histogram
shows the photometric redshift distribution of the five $J_s -K_s$
selected galaxies without spectroscopic redshift. Two
are in the range
$1.7<z_{\rm phot}<2$,  and three have $z_{\rm phot}>2$. The median
is $2.4$, similar to the median photometric redshift of
galaxies with spectroscopic redshifts ($2.6$).  We conclude
that there is no evidence for a strong
redshift bias in our spectroscopic sample.

The redshifts of the five confirmed $z>2$ galaxies span a large
range, which is not surprising given the large spacing between
the $J_s$ and $K_s$ filters. The mean redshift
is 2.7, similar to the median photometric
redshift of 2.6 for galaxies with $J_s- K_s>2.3$ and $K_s<22.5$ in \hdfs\
(Franx et al.\ 2003). Remarkably, three
galaxies are at almost identical
redshift, qualitatively consistent with the strong angular clustering
of $J_s-K_s$ selected galaxies in \hdfs\ (Daddi et al.\ 2003).

The $J_s - K_s$ selection is the higher
redshift analog of the well known selection of $1 < z < 2$ galaxies by
their $I - H$ color (e.g., {McCarthy} {et~al.} 2001), and by combining the
two methods galaxies can be selected over the entire
range $1.0 \lesssim z \lesssim 3.5$ (see Fig.\ \ref{zdist.plot}).
We note that
$I - H$ selected galaxies appear to
be highly clustered as well: only one out
of twelve objects is {\em not} within 2000\,\kms\ of another galaxy.

\null
\vbox{
\begin{center}
\leavevmode
\hbox{%
\epsfxsize=7.8cm
\epsffile{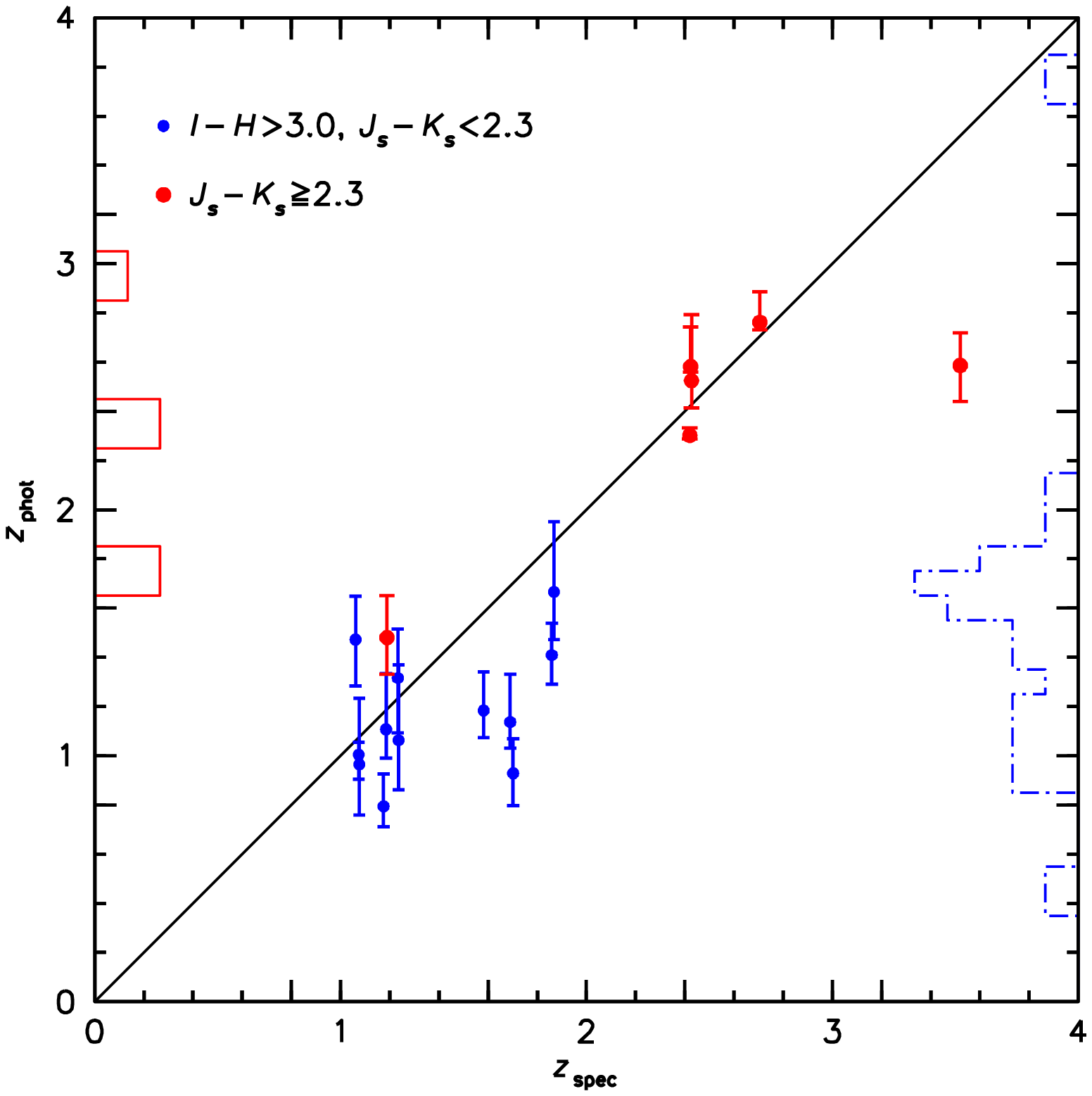}}
\figcaption{\small
Spectroscopic and photometric redshifts of
galaxies selected by broadband colors. The
twelve galaxies with $I - H>3.0$ and $J_s - K_s < 2.3$
have redshifts in the range $1.06- 1.87$. Five out of six
galaxies with $J_s - K_s > 2.3$ are at $2.43 \leq z \leq 3.52$.
Histograms show the photometric redshifts of observed
galaxies whose spectra had insufficient S/N to
determine their spectroscopic redshifts.
\label{zdist.plot}}
\end{center}}

\begin{small}
\begin{center}
{ {\sc TABLE 1} \\
\sc $J_s-K_s$ Selected Galaxies with Redshifts} \\
\vspace{0.1cm}
\begin{tabular}{lccccc}
\hline
\hline
Galaxy & $K_s$ & $K_{s,\,\rm corr}^a$ & $J_s-K_s^b$
& $z$ & $M_{V,{\rm rest}}^c$ \\
\hline
596 & 19.68 & 19.91 & 2.35 &  $1.189$ & --- \\
1671 & 19.09 & 19.23 & 2.65 & $2.424$ & $-23.7$ \\
1195 & 19.29 & 19.63 & 2.30 & $2.425$& $-23.9$ \\
1458 & 19.71 & 19.86 & 2.25 & $2.427$ & $-23.2$ \\
184 & 19.29 & 19.40 & 2.54 &  $2.705$ & $-24.1$ \\
1656 & 19.59 & 19.81 & 2.98 & $3.525$ & $-24.8$ \\
\hline
\end{tabular}
\end{center}
{\small{
$^a$\,$K$-band magnitudes corrected for weak lensing amplification
by the foreground $z=0.83$ galaxy cluster ({Hoekstra}, {Franx}, \&  {Kuijken} 2000).\\
$^b$\,Colors were recalibrated following the initial spectroscopic
selection; errors are $\approx 0.05$ magnitudes.\\
$^c$\,$H_0 = 65$\,\kms\,Mpc$^{-1}$, $\Omega_m=0.3$, and
$\Omega_{\Lambda}=0.7$.
}}
\end{small}
\\

\section{Spectra of Red $z>2$ Galaxies}

\begin{figure*}[t]
\begin{center}
\leavevmode
\hbox{%
\epsfxsize=14.2cm
\epsffile{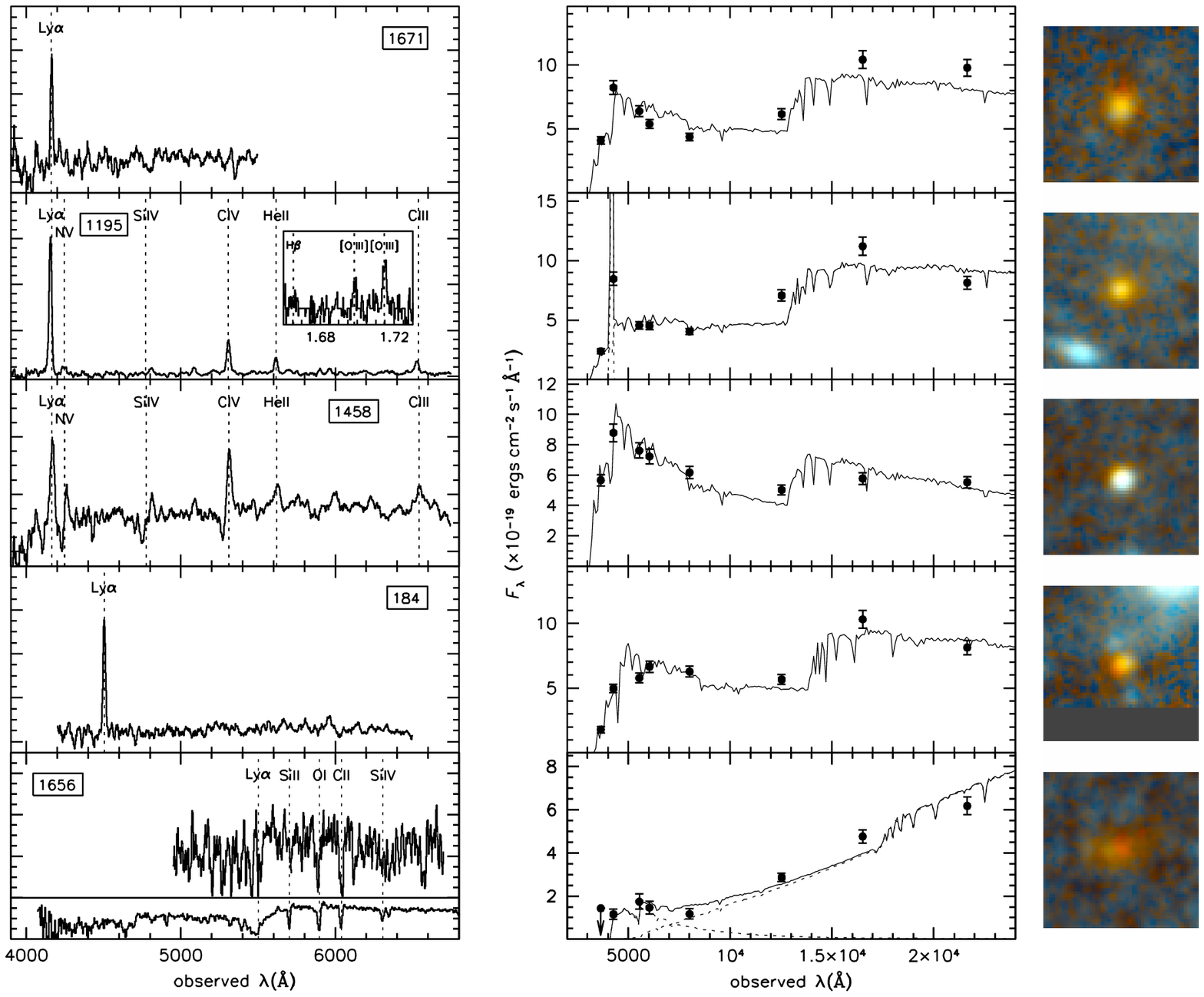}}
\figcaption{
\small
{\em Left panels:} Rest-frame UV spectra of galaxies at $z>2$.
The spectra display a wide range of properties. The inset for galaxy 1195 shows
a NIRSPEC rest-frame optical spectrum. The panel
below 1656 shows the mean spectrum of $\sim
250$ Lyman-break
galaxies without Ly$\alpha$ emission, from Shapley et al.\ (2002, in
preparation). {\em Right panels:} Spectral energy distributions of the same
galaxies. Overplotted are models with varying age and dust content
(see text). The $B$,$H$, and probably $J_s$ band fluxes
of 1195 are affected by strong emission lines.
Color images were created from the {\em HST} $R_{AB}\equiv
0.5\times(V_{\rm F606W}^{\rm AB}+I_{\rm F814W}^{\rm AB})$
and VLT $K_s$ images (after smoothing to the same resolution),
and are $5\farcs 1 \times 5\farcs 1$.
\label{specseds.plot}}
\end{center}
\end{figure*}

The rest-frame UV spectra of confirmed $z > 2$ galaxies are shown in
Fig.\ \ref{specseds.plot}.
The small sample displays a wide range of properties: two galaxies
have emission lines characteristic of Active Galactic Nuclei (AGN),
two are Ly$\alpha$ emitters with no evidence for the presence of
an AGN, and one shows
interstellar absorption
lines typical of nearby star-forming galaxies.

The high fraction of emission line galaxies is striking, although
in part a selection effect resulting from the difficulty of measuring
absorption line redshifts of these faint galaxies. Our single absorption
redshift required an exposure time of 72\,ks, and judging from
a cross-correlation with
a LBG template it is only marginally significant ($r=3.1$).
Two other galaxies
show evidence for breaks indicating the onset of the
Ly$\alpha$ forest, but do not produce significant peaks in their
cross-correlation functions. Since there are five observed galaxies
without redshift we estimate that the true emission line
fraction is $\gtrsim 40$\,\%. The fact that Ly$\alpha$ photons
can escape in some $J_s-K_s$ selected galaxies may indicate that they
are not all very dusty, or that they have
patchy dust distributions.


The two AGN are at
identical redshift and separated by 84\arcsec.  Galaxy 1195 is a
narrow-line AGN similar to Type-II
quasars and nearby Seyfert-II galaxies.
Galaxy 1458 has a complex spectrum, with
narrow ($\sim 1500$\,\kms\ FWHM) Ly$\alpha$ and C\,{\sc iv} emission
lines and broader ($\sim 3000$\,\kms) He\,{\sc ii} and C\,{\sc iii}
lines. The resonant transitions N\,{\sc v}, Si\,{\sc iv}, and C\,{\sc iv}
show P-Cygni profiles, indicative of a radiatively driven $\sim
3000$\,\kms\ wind (e.g., {Murray} {et~al.} 1995).
%
More data are needed to determine how common AGN are among
$J_s-K_s$ selected galaxies, and to compare the rate of occurrence
to that in rest-UV selected objects ({Steidel} {et~al.} 2002).
It is widely believed that all massive,
bulge-dominated systems in the local universe
harbor massive black holes (e.g., {Magorrian} {et~al.} 1998) and we
can speculate that the optical-break galaxies are progenitors of
such galaxies observed at a time when many black holes were still
in the process of formation.

\section{Spectral Energy Distributions}

Broad band spectral energy distributions (SEDs) are shown in the
central panels of Fig.\
\ref{specseds.plot}. Optical data were obtained with the {\em Hubble
Space Telescope} ({\em HST}) WFPC2
({van Dokkum} {et~al.} 2000) and VLT+FORS; a full description
of the photometry is given in N.\ M.\ F\"orster Schreiber et al.\ (in
preparation). We fitted GISSEL98 model spectra from
Bruzual \& Charlot (in preparation) to interpret the SEDs.
As is well known it is difficult to
obtain unique solutions for the functional form of the star formation
history, the age of the stellar population, and the dust content (see,
e.g., {Papovich} {et~al.} 2001; {Shapley} {et~al.} 2001). Here we only
consider models with constant star formation rate and a
{Calzetti} {et~al.} (2000) reddening law.

As shown in Fig.\ \ref{specseds.plot}
the two Ly$\alpha$ emitting galaxies 1671 and 184 are fairly
well fitted by such models. The same model
with a ``maximally old'' age of 2.6\,Gyr and moderate
extinction $E(B-V)=0.4$
provides the best fit to both galaxies, and we conclude
that the red $J_s-K_s$ colors of these objects are most likely the result
of their evolved stellar populations.
The SEDs of galaxies 1195 and 1458 are almost certainly affected by
emission from their AGN, which is difficult to model.
When the continuum contributions of the AGN
are ignored, galaxy 1458 is best fitted by
a constant star formation model of age 1.4\,Gyr and $E(B-V)=0.3$,
and 1195 by a model with age 1.7\,Gyr and $E(B-V)=0.5$.
The highest redshift galaxy in our sample, 1656, is not well fitted by
simple models. Good fits are obtained for
composite models of a young, dust-free component and a second young
component which is heavily obscured, having $E(B-V)\sim 1.5$
(see Fig.\ \ref{specseds.plot}).

For galaxy 1195 the contribution of
Ly$\alpha$ emission to the $B$-band flux (30--40\,\%) had to be taken
into account in the model fitting.
On January 6
2002 we obtained a 1200 s $H$-band spectrum with NIRSPEC on Keck II
(inset in Fig.\ \ref{specseds.plot}).
The spectrum shows the redshifted [O\,{\sc iii}] $\lambda$\,4959,5007
lines; H$\beta$ is undetected. The lines contribute $0.12 \pm 0.04$
magnitudes to the $H$-band flux, consistent with the observed offset
from the best fitting model spectrum.
Since 1195 is by far the strongest
Ly$\alpha$ emitter in our sample this result
suggests that the contribution of line emission to the NIR fluxes
is generally small for $J_s-K_s$ selected galaxies.

We conclude that the red $J_s-K_s$ colors are probably produced
by evolved stellar populations and in one case a dusty young population.
The median age of 1.7\,Gyr is higher than the
median age of $\sim 3 \times 10^8$\,yr of
LBGs, derived using the same models
({Shapley} {et~al.} 2001; {Papovich} {et~al.} 2001).

\section{Discussion}

The spectroscopic redshifts
demonstrate that $z\gtrsim 2$ galaxies can be selected
efficiently by  $J_s-K_s$ colors alone. The criterion
$J_s-K_s>2.3$ is not extreme, but selects galaxies whose
rest-frame optical colors are
similar to those of normal nearby galaxies:
for $z = 2.7$ our $J_s - K_s$ limit corresponds to $U
- V \gtrsim 0.1$ in the rest-frame, and a local sample selected thus would
include almost all luminous galaxies (e.g., {Jansen} {et~al.} 2000).

The surface density of $J_s-K_s>2.3$ galaxies in the \ms1054\ field is
$1.09^{+0.20}_{-0.16}$\,arcmin$^{-2}$ to $K=21$.  If our small
spectroscopic sample is representative for the general population,
$\sim 20$\,\% are at low redshift, and the surface
density of $z>2$ optical-break galaxies is $\approx
0.9$\,arcmin$^{-2}$ (ignoring the lensing effect of the foreground
cluster).  The Poissonian errors imply 65\,\% confidence limits of 0.7
-- 1.1. However, significant field-to-field variations exist: there
are no $J-K>2.3$ galaxies to $K=21$ in HDF-N ({Fern{\' a}ndez-Soto},
{Lanzetta}, \& {Yahil} 1999), whereas the surface density in HDF-S is
similar to that in the $\sim 5 \times$ larger \ms1054\ area.
Additional errors due to galaxy correlations increase the upper
limits, but leave the lower limits relatively unchanged (see, e.g.,
Daddi et al.\ 2000). Based on the observed clustering of $J_s-K_s>2.3$
galaxies in \hdfs\ (Daddi et al.\ 2003), we find a 2$\sigma$ lower
limit of $\approx 0.5$\,arcmin$^{-2}$.

The surface density can be compared to that
of LBGs with $K_s\leq 21$,
by integrating the $K_s$-band luminosity function determined
for $z \approx 3$ LBGs
by {Shapley} {et~al.} (2001). We assume that the luminosity function does
not evolve, and use a
template spectrum
to calculate the dependence of $M_K^*$
on redshift. Integrating over the redshift range $2.0-3.5$
we find a surface density of $\approx 2.0$~arcmin$^{-2}$
to $K_s=21$, twice the density of optical-break
galaxies. This result is quite
sensitive to the assumption that the redshift distribution of
optical-break galaxies is
a tophat with a lower bound of $2.0$.
Integrating over the redshift range $2.4-3.5$ reduces the
density to $\approx 0.9$\,arcmin$^{-2}$, and we conclude
that the surface density of optical-break selected galaxies is $0.5-1 \times$
that of Lyman-break selected galaxies in the same redshift range.
These numbers are broadly
consistent with results obtained for fainter samples
in \hdfs\ (Franx et al.\ 2003).

Optical-break galaxies are underrepresented in samples selected by
the Lyman-break technique:
only two out of 40 galaxies observed in $J$ and $K_s$
by {Shapley} {et~al.} (2001) satisfy our selection criterion. 
Most optical-break galaxies in the \ms1054\ field
are too faint in the rest-frame UV to be
selected as LBGs in ground-based surveys:
$\sim 70$\,\%  of galaxies with $K_s<21$
have $R_{\rm AB} > 25.5$. As shown in Franx et al.\ (2003)
the overlap is even smaller for the fainter \hdfs\ sample.
We note however that -- because of obvious selection
effects -- four of the five galaxies confirmed by optical spectroscopy
are brighter than
this limit, and probably would be selected as LBGs if at the appropriate
redshift.



The most straightforward (but not unique) interpretation of the newly
identified population is that they are evolved descendants of
galaxies that started forming stars at redshifts $z>4$.
As shown in Franx et al.\ (2003) their typical mass-to-light ratios
are probably higher
than those of galaxies selected by the Lyman-break technique;
hence they may be the most massive galaxies
yet identified at $z>2$, and could be progenitors of early-type
galaxies and bulges.
Interestingly, 
the five confirmed $z>2$ galaxies alone may already place limits
on galaxy formation models. They constitute $\approx 1.5$\,\% of the
$K_s<20$ counts (e.g., {Maihara} {et~al.} 2001, Labb\'e et al.\ 2002), whereas
current semianalytical galaxy formation models have predicted that
galaxies beyond $z\sim 2$ should essentially be absent in such bright samples
(see {Cimatti} {et~al.} 2002).
The fraction is even higher ($\sim 5$\,\%) when
galaxies without spectroscopic redshift are included.

Surveys over larger areas
will help establish the surface density with greater confidence.
More spectroscopy is needed to determine the redshift distribution
and AGN fraction better.
Finally, NIR spectroscopy for more galaxies
is needed to determine the contribution of emission lines
to the broad band fluxes.

\acknowledgements{
We thank Henk Hoekstra for computing the lens magnifications,
Alice Shapley and Aaron Barth for useful discussions, and the
referee, Pat McCarthy, for constructive comments which improved
the paper. P.~G.~v.~D.\ acknowledges
support by NASA throught the SIRTF Fellowship program.
The authors wish to extend
special thanks to those of Hawaiian ancestry on whose sacred mountain
we are privileged to be guests. Without their generous hospitality,
many of the observations presented herein would not have been possible.
}


\end{document}